\documentclass[conference]{IEEEtran}

\usepackage{xcolor} 

\usepackage{tabularx}
\usepackage{booktabs}
\usepackage{longtable}
\usepackage{multirow}
\usepackage{colortbl}
\usepackage{hhline}
\usepackage{tabularx,colortbl}

\usepackage{hyperref} 
\usepackage{cite}
\usepackage{amsmath,amssymb,amsfonts}
\usepackage{mathrsfs} 
\usepackage{algorithmic}
\usepackage{graphicx}
\usepackage{enumitem}
\usepackage{soul}

\usepackage{subfig}
\usepackage{mathtools} 

\usepackage{color}

\usepackage{tikz}

\makeatletter
\newcommand\notsotiny{\@setfontsize\notsotiny\@vipt\@viipt}
\makeatother

\makeatletter
\newcommand*\bigcdot{\mathpalette\bigcdot@{.5}}
\newcommand*\bigcdot@[2]{\mathbin{\vcenter{\hbox{\scalebox{#2}{$\m@th#1\bullet$}}}}}

\newcommand{\mathbolditalic}[1]{\text{\textit{\textbf{#1}}}}
\newcommand{\etal}{\textit{et al. }}
\makeatother

\usepackage{textcomp}
\def\BibTeX{{\rm B\kern-.05em{\sc i\kern-.025em b}\kern-.08em
    T\kern-.1667em\lower.7ex\hbox{E}\kern-.125emX}}

\DeclareRobustCommand*{\IEEEauthorrefmark}[1]{%
  \raisebox{0pt}[0pt][0pt]{\textsuperscript{\footnotesize #1}}%
}

\definecolor{g}{rgb}{0.56, 0.93, 0.56}

\begin{document}

\title{A Multi-Authority Attribute-Based Signcryption Scheme with Efficient Revocation for Smart Grid Downlink Communication}
	
\author{
    \IEEEauthorblockN{
        Ahmad Alsharif\IEEEauthorrefmark{1,2},
        Ahmad Shafee\IEEEauthorrefmark{2},
        Mahmoud Nabil\IEEEauthorrefmark{2},
        Mohamed Mahmoud\IEEEauthorrefmark{2}, and
        Waleed S. Alasmary\IEEEauthorrefmark{3}
	}
    \newline\IEEEauthorblockA{\IEEEauthorrefmark{1}Department of Computer Science, University of Central Arkansas, AR, USA 72035.}
    \newline\IEEEauthorblockA{\IEEEauthorrefmark{2}Department of Electrical and Computer Engineering, Tennessee Tech. University, TN, USA 38505.}
    \newline\IEEEauthorblockA{\IEEEauthorrefmark{3} Computer Engineering Department, Umm Al-Qura University, Makkah 21421, Saudi Arabia.}
}
	
\maketitle

\begin{abstract}

In this paper, we propose a multi-authority attribute-based signcryption scheme with efficient revocation for smart grid downlink communications.
In the proposed scheme, grid operators and electricity vendors can send multicast messages securely to different groups of consumers which is required in different applications such as firmware update distribution and sending direct load control messages.
Our scheme can ensure the confidentiality and the integrity of the multicasted messages, allows consumers to authenticate the source of the multicasted messages, achieves and non-repudiation property, and allows prompt revocation, simultaneously which are required for the smart grid downlink communications.
Our security analysis demonstrates that the proposed scheme can thwart various security threats to the smart grid.
Our experiments conducted on an advanced metering infrastructure (AMI) testbed confirm that the proposed scheme has low computational overhead.

\end{abstract}

\IEEEpeerreviewmaketitle

\section{Introduction}

The smart grid (SG) is the next generation of the traditional power grid.
It integrates information and communication technologies with traditional power grid to provide two-way communications between the grid’s major entities including grid operators, electricity vendors, and electricity consumers to ensure the efficient and reliable operation of the grid.
One of the main components of the SG is the advanced metering infrastructure (AMI) networks which
connect smart meters (SMs) installed at consumers' houses to the grid operators and vendors.

\begin{figure*}[!t]
	\centering
	\includegraphics[clip=true,width=0.9 \textwidth]{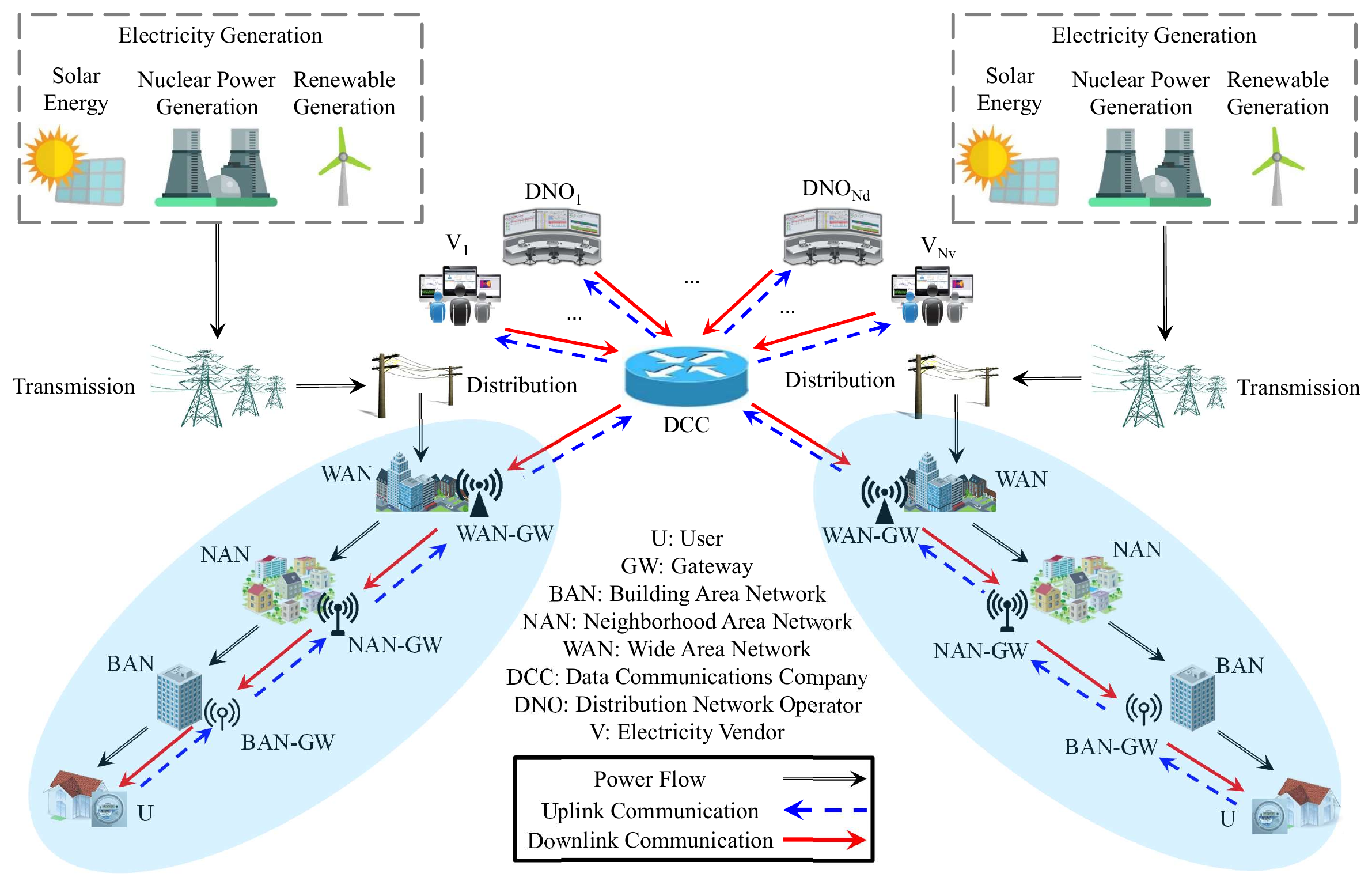}
	\caption{Network model for SG downlink communications.}
	\label{fig:multicast_network_model}
    \vspace{-5mm}
\end{figure*}

Multi-authority AMI networks, which are deployed in most European countries  and several states in the U.S., allow energy deregulation, i.e., electricity retailing through different electricity vendors \cite{EU, DEP2SA, EPDA}.
Therefore, consumers not only can choose from a number of independent third party electricity vendors, but pricing options are also more plentiful due to the competition between these different vendors.
\autoref{fig:multicast_network_model} shows the conceptual architecture a multi-authority AMI network \cite{EPDA}.
As shown in the figure, data communication  can be either uplink or downlink communication.

In the uplink data communication, data is sent by SMs to grid operators and vendors.
This can allow the automated collection of metering data in which grid operators and vendors collect fine-grained power consumption data (PCD) at high rates, e.g., few minutes, for real-time grid monitoring, and energy distribution management.
For example, fine-grained data analysis can be used for the reduction of the peak-to-average ratio which can help in preventing
blackouts, failures to supply electricity \cite{managment1,managment2,report_june_2017}.
Also, fine-grained PCD are needed for real-time price-based demand/response programs in which electricity prices vary depending on the supply-to-demand ratio especially during peak hours \cite{DR1,DR2}.

On the other hand, in the downlink data communications, data is sent by grid operators or electricity vendors to a group of SMs.
The downlink communication should ensure secure multicast for different applications.
For example, sending firmware and configuration updates to a group of SMs in specific areas requires secure multicast \cite{baza2018blockchain,tonyali2017attribute2}.
Also, in direct load control (DLC) demand/response programs, grid operators need to send DLC messages to a group of users, that subscribe to the same DLC demand/response plan, in order to turn off/on some specific load during peak/regular hours \cite{roy2014lte,saxena2015exploiting}.
Moreover, electricity vendors may send charging schedules to a group of users in selected areas to charge their electric vehicles or home batteries \cite{nabil2019priority,pazos2018secure}.
Furthermore, electricity vendors may send energy trading requests to a group of users in selected areas that are subscribed to energy charging/discharging plans, asking for energy injection to the SG during peak hours \cite{sherif2018privacy,baza2018blockchain2}.

Extensive research has been conducted to study the security and privacy issues in AMI networks.
However, most of the existing schemes address consumers' privacy, data integrity and authenticity
in uplink communication \cite{survey1}.
Few schemes have been proposed to study     the security of the downlink communications in AMI networks.
Moreover, the IEEE 802.11 protocol, which is the underlying protocol for AMI networks cannot be used for secure multicast communication efficiently and effectively \cite{tonyali2017attribute}.
Therefore, there is a need for a scheme that not only allows secure multicast communication, but also considers the unique characteristics of multi-authority AMI networks.
Specifically, a good secure multicast scheme should ensure message confidentiality, i.e., the multicasted messages can be decrypted only by the intended users.
In addition, the scheme should allow dynamic group memberships, i.e., members' enrollment/revocation should be done efficiently and promptly since they can subscribe/unsubscribe to any plan with any vendor at any time.
Moreover, users should be able to authenticate the senders of the multicasted messages.
Furthermore, message non-repudiation property should be achieved.

In order to address the aforementioned challenges, we propose in this paper a multi-authority attribute-based signcryption scheme that can be used to secure SG downlink communications.
We construct our scheme based on, but not limited to, the multi-authority attribute based encryption (MA-ABE) scheme proposed in \cite{RW2015_MA_ABE}.
To the best of our knowledge, this paper proposes the first fully decentralized multi-authority attribute based signcryption scheme that can ensure data confidentiality, sender authentication, and non-repudiation, and allow prompt attribute revocation, simultaneously.

The remainder of this paper is organized as follows.
Related works are discussed in Section \ref{sec:multicast_related_works}.
The considered system models and the design goals are presented in Section \ref{sec:multicast_model_requirements}.
Preliminaries are given in Section \ref{sec:multicast_preliminaries}.
The proposed scheme is explained in Section \ref{sec:multicast_scheme}.
The security analysis and performance evaluation are given in Sections \ref{sec:multicast_security_analysis} and \ref{sec:multicast_performance}, respectively.
Conclusions are drawn in Section \ref{sec:multicast_conclusions}.

\section{Related Works} \label{sec:multicast_related_works}

The SG downlink communication has been considered in several schemes  \cite{ye2015hibass,alharbi2016efficient,baza2015efficient}.
However, these schemes consider only broadcast downlink communication and cannot support multicast communication.
Several schemes  have been proposed to ensure fine-grained access control and/or secure multicast for the SG communications \cite{liu2014achieving,fadlullah2012toward,ABSC2015}.
In the ABE scheme proposed in \cite{LW2011_MA_ABE}, Liu \etal proposed a multi-authority access control scheme with attribute revocation for the SG \cite{liu2014achieving}. In the proposed scheme, if a user's attributes can satisfy the access policy associated with a ciphertext, this user can decrypt that ciphertext only after receiving a unique token from a central entity called third party auditor (TPA).
Therefore, the proposed scheme \cite{liu2014achieving} cannot support multicast communication efficiently since the TPA must send a unique token to each member in a multicast group using unicast communication, i.e., a unicast downlink communication is needed to decrypt any multicast message.
In \cite{fadlullah2012toward}, Fadlullah \etal have proposed a secure multicast scheme for SG communications using the key-policy attribute-based encryption (KP-ABE) \cite{ABE2}. However, the scheme is limited to single attribute authority, i.e, a single authority controls all attributes. Also, it does not support sender authentication and non-repudiation.
In \cite{ABSC2015}, Hu \etal have proposed an attribute-based signcryption scheme to secure multicast communications in the SG.
The scheme proposes a modification to CP-ABE \cite{ABE3} in order to achieve attribute-based encryption, data-origin authenticity and non-repudiation.
However, the scheme is limited to single attribute authority and does not support attribute revocation.

Different from the above schemes, our scheme allows
(1) multiple authorities to issue and control their own attributes;
(2) data-origin authenticity and non-repudiation;
and
(3) prompt attribute revocation, simultaneously.

\section{System Models and Design Goals} \label{sec:multicast_model_requirements}

\subsection{Network Model}
The considered network model is shown in \autoref{fig:multicast_network_model}.
This model was used in \cite{DEP2SA, EPDA} to secure uplink smart grid communications. In this paper, we aim to secure the downlink communications.
The network model has the following entities.
\begin{itemize}

    \item \textit{Distribution Network Operators} (DNOs). We consider a set of DNO companies, $\mathbb{D}=\{D_j, 1 \leq j \leq N_d\}$. Each $D_j$ is licensed to distribute electricity in a particular geographic area $j$. Each DNO manages and operates the distribution networks within its area.
	
    \item \textit{Electricity Vendors}. We consider a set of electricity vendor companies, $\mathbb{V}=\{V_k, 1 \leq k \leq N_s\}$. Each $V_k$ is responsible for supplying electricity to its users who may be located at different areas.
	
    \item \textit{Users}. We consider a set of users $\mathbb{U}=\{u_i, 1 \leq i \leq N_u\}$.

        Users can change from one vendor to another at any time.
        In addition, users can add, change, and remove plans offered by the same vendor at any time.
        An SM is installed at each user's house that communicates with the DNOs and electricity vendors through a node called the data communication company (DCC).
	
    \item \textit{Data Communication Company} (DCC). It has the responsibility of delivering the downlink communications received from operators and vendors to users.
	
    \item \textit{Networking Facilities}. They form a hierarchical network structure to connect the DCC to SMs at users' side through a WAN-GW, a NAN-GW, and a BAN-GW as shown in the figure.
\end{itemize}

\subsection{Threat Model}

There exist an adversary $\mathscr{A}$ that can eavesdrop all the transmitted messages.
$\mathscr{A}$ may try to decrypt the multicasted messages to revel any sensitive information sent to any group of users.
Users also may try to breach data confidentiality, i.e, they may try to decrypt the multicasted messages intended to other groups of users.
In addition, a malicious user may collude with other users or $\mathscr{A}$ in order to decrypt a ciphertext that they can not decrypt individually.
Moreover, $\mathscr{A}$ may try to launch active attacks by injecting malicious messages to any group of users, e.g. sending malwares instead of firmware updates to have full control on their devices.

\subsection{Design Goals}

Based on the aforementioned network and threat models, the following goals should be achieved.
\begin{enumerate}
  \item \textit{Secure multicast and data confidentiality}. Only selected users by the grid operators or vendors should be able to decrypt the multicasted messages. Other users should be prevented from accessing these messages.
  \item \textit{Collusion resistance}. Users that are not supposed to decrypt a ciphertext individually, should not be able to decrypt it even if they collude together by using their secret keys.
  \item \textit{Sender authentication and non-repudiation}. Users should be able to authenticate the sender of the multicasted messages. Messages form $\mathscr{A}$ should be detected and discarded. Also, non-repudiation property should be achieved.
      \item \textit{Prompt Revocation}. Only valid, i.e. non-revoked, users should be able to decrypt the multicasted ciphertext. Revocation process should be done immediately without any delays.
\end{enumerate}

\section{Preliminaries} \label{sec:multicast_preliminaries}

\subsection{The Chinese Remainder Theorem}

Let $\{q_1, q_2, \dots, q_m\}$ be $m$ pairwise relatively primes and let $\{b_1, b_2, \dots, b_m\}$ be $m$ arbitrary integers. The Chinese Remainder Theorem (CRT) states that the system of congruences $ B \equiv b_i \mod q_i \text{ for } 1 \leq i \leq m $ has a unique solution modulo $Q= \prod_{i=1}^{n} q_i$. The unique solution $B$ is given by
\begin{equation*}
    B=\sum_{i=1}^{m} b_iQ_iy_i \mod Q
\end{equation*}
where $Q_i=\frac{Q}{q_i}$ and $y_i \equiv \frac{1}{Q_i} \mod q_i$ for  $1 \leq i \leq m$.

\subsection{Multi-Authority Attribute-Based Encryption \cite{RW2015_MA_ABE}}
Let $\mathcal{U}=\{u_1, \dots, u_n\}$ be the universe of the attributes, $\mathcal{U_\theta}=\{\theta_1, \dots, \theta_n\}$ be the universe of the attribute authorities controlling the $n$ attributes, and $\mathcal{GID}=\{\text{GID}_1, \dots, \text{GID}_m\}$ be the universe of the global identities that identify $m$ users.

\subsubsection{Linear Secret Sharing and Access Policy}
We use the same definition for linear secret sharing (LSS) and access policy as in \cite{RW2015_MA_ABE} and \cite{LW2011_MA_ABE}.
Any monotonic boolean formula over $\mathcal{U}_\Theta$ can be represented as an access matrix as follows.
Let $p$ be a prime.
A secret-sharing scheme $(\Pi)$ over a set of attributes $\mathcal{U}$ is called linear (over $\mathbb{Z}_p$) if

\begin{enumerate}
    \item The $\ell$ shares of a secret $z \in \mathbb{Z}_p$ for each attribute form a vector $\boldsymbol \lambda$ over $\mathbb{Z}_p$.

    \item There exists a matrix $A$ called the share-generating matrix for $\Pi$. The matrix $A$ has $\ell$ rows and $n$ columns. For all $ 1 \leq x \leq \ell$, the $x^{th}$ row of $A$ is labeled by an attribute $\delta(x)$, where $\delta$ is a function that maps rows of $A$ to attributes from $\mathcal{U}$, i.e., $\delta$$: \{1,\dots, \ell\} \rightarrow \mathcal{U}$.
        When we consider the column vector $\mathbolditalic{v} = (z, r_2, \dots , r_n)$, where
        $\{r_2, \dots, r_n\} \xleftarrow[]{R} \mathbb{Z}_p$ are randomly chosen, then $\boldsymbol \lambda = A\mathbolditalic{v}$ is the vector of $\ell$ shares of the secret $z$ according to $\Pi$. The share $\lambda_x$ belongs to the attribute $\delta(x)$.

\end{enumerate}
As mentioned in \cite{RW2015_MA_ABE,LW2011_MA_ABE}, each secret-sharing scheme should satisfy the following requirements:
\begin{itemize}
  \item A reconstruction requirement, i.e., each authorized set of attributes can reconstruct the secret.
  \item A security requirement, i.e., other sets of attributes, unauthorized sets,  cannot reveal any information about the secret.
\end{itemize}

For example, let S denote an authorized set of attributes and let I be the set of rows whose labels are in S. There exist constants $\{c_i\}_{i \in I} \in \mathbb{Z}_p$ such that for any valid shares
of a secret $z$ according to $\Pi$, it is true that: $\sum_{i \in I} c_i \lambda_i=z$, or equivalently $\sum_{i \in I} c_i \mathbolditalic{A}_i=(1, 0, \dots, 0)$, where $\mathbolditalic{A}_i$ is the $i^{th}$ row of $A$.
In Appendix \ref{appx_A}, we give an example of generating the access matrix, computing the vector of secret shares $\boldsymbol \lambda$ and the reconstruction coefficients $\{c_i\}_{i\in I}$ from a boolean formula.

\subsubsection{Algorithms}
The scheme in \cite{RW2015_MA_ABE} consists of the following algorithms.

\begin{itemize}

\item $\mathsf{GlobalSetup}(1^\kappa)\rightarrow \text{GP}$

    This algorithm takes a security parameter $\kappa$ and outputs the public global parameters for the system.  The global parameters (GP) includes $\mathcal{U}$, $\mathcal{U}_\Theta$, $\mathcal{GID}$, and $\mathsf{T}$ which is a mapping function that maps each attribute in $\mathcal{U}$ to a unique authority in $\mathcal{U}_\Theta$, i.e., $\mathsf{T}$$: \mathcal{U}\rightarrow \mathcal{U}_\Theta$.

\item $\mathsf{AuthoritySetup}(\text{GP}, \theta)\rightarrow \text{PK}_\theta,\text{SK}_\theta$

This algorithm generates a public/private key pair $\{\text{PK}_\theta,\text{SK}_\theta\}$ for each attribute authority $\theta \in \mathcal{U}_\Theta$.

\item $\mathsf{KeyGen}(\text{GID}, \theta , u, \text{SK}_\theta , GP) \rightarrow \text{SK}_{\text{GID}, u}$

This algorithm takes the global identity of a user $\text{GID} \in \mathcal{GID}$, an attribute $u \in \mathcal{U}$, the authority $\theta$ controlling the attribute $u$, the secret key of the authority $\text{SK}_\theta$, and the global parameters  $\text{GP}$. The algorithm outputs $\text{SK}_{\text{GID}, u}$ which is a secret key for the identity-attribute pair used for decryption.

\item $\mathsf{Encrypt}(M, (A,\delta) , \{\text{PK}_\theta\}, \text{GP}) \rightarrow \text{CT}$\\
This algorithm takes a message $M$, an access policy $(A,\delta)$, a set of public keys $\{\text{PK}_\theta\}$ of the authorities controlling attributes in the access policy, and the global parameters $\text{GP}$. The algorithm outputs the ciphertext $\text{CT}$.

\item $\mathsf{Decrypt}(\text{CT}, \{\text{SK}_{\text{GID}, u}\}, \text{GP}) \rightarrow M \text{ or } \perp$\\
This algorithm takes the ciphertext $\text{CT}$, the set of secret keys $\{\text{SK}_{\text{GID}, u}\}$ of a single user with identity $\text{GID}$ corresponding to different attributes, and the global parameters, and outputs $M$ if and only if the attribute set associated with $\{\text{SK}_{\text{GID}, u}\}$ can satisfy the access policy of the ciphertext, otherwise, decryption fails.

\end{itemize}

\section{The proposed scheme} \label{sec:multicast_scheme}
In this section, we first provide the construction of our attribute-based signcryption scheme. Then, we discuss how the scheme can be applied to secure the SG downlink communications.

\subsection{Definitions}

Let
$\mathcal{U}_1=\{u_1, \dots, u_j\}$ be the universe of $j$ attributes,
$\mathcal{U}_\Theta=\{\theta_1, \dots, \theta_j\}$ be the universe of the attribute authorities controlling the $j$ attributes,
$\mathcal{U}_\Phi=\{\phi_1, \dots, \phi_k\}$ be the universe of entities allowed to signcrypt messages,
$\mathcal{U}_2=\{s_1, \dots, s_k\}$ be the universe of $k$ identity attributes corresponding to $k$ signers,
$\mathcal{GID}=\{\text{GID}_1, \dots, \text{GID}_m\}$ be the universe of the global identities that identifies $m$ users, and
$\mathcal{Q}=\{q_1, \dots, q_m\}$ be $m$ pairwise relatively prime positive integers where each prime $q_i$ is assigned to a user with $\text{GID}_i$.
Let $\mathcal{U} = \mathcal{U}_1 \cup \mathcal{U}_2$ and $\mathcal{U}_\Psi = \mathcal{U}_\Theta \cup \mathcal{U}_\Phi$.

In addition, we use the same access structure as \cite{RW2015_MA_ABE,LW2011_MA_ABE} with an additional restriction.
The access structure encoded as a monotonic boolean formula over $\mathcal{U}_\Theta$ and $\mathcal{U}_\Phi$ should be on the form ``The signer identity attribute $s \in \mathcal{U}_2$'' \textbf{AND} ``any monotonic boolean formula over $\mathcal{U}_1$''.
Therefore, in order to designcrypt a signcryppted text, a user should do the following
\begin{itemize}
  \item Uses the verification key corresponding to the signer $\phi$ controlling the signer identity attribute $s$.
  \item Possess attributes satisfying the second part of the boolean formula.
\end{itemize}

Moreover, let $G_x \subset \mathcal{GID}$ be the set of users who holds an attribute $u_x$. We refer to $G_x$ as the access list of the attribute $u_x$. Let $\mathcal{G}=\{G_1, \dots, G_j\}$ be the universe of access lists of the $j$ attributes defined in $\mathcal{U}_1$.

\subsection{Algorithms}
Our scheme consists of the following eight algorithms.

\begin{enumerate}

\item $\mathsf{GlobalSetup}(1^\kappa)\rightarrow \text{GP}$.\\
    This algorithm takes a security parameter $\kappa$ and outputs the public global parameters for the system.  The global parameters (GP) includes $\mathcal{U}$, $\mathcal{U}_\Psi$, $\mathcal{GID}$, and $\mathsf{T}$ which is a mapping function that maps each element in $\mathcal{U}$ to a unique element in $\mathcal{U}_\Psi$, i.e., $\mathsf{T}$$: \mathcal{U}\rightarrow \mathcal{U}_\Psi$. More specifically,
    $\mathsf{T}$ maps each element in $\mathcal{U}_1$ to a unique element in $\mathcal{U}_\Theta$ and maps each element in $\mathcal{U}_2$ to a unique element in $\mathcal{U}_\Phi$.

\item $\mathsf{SignKeyGen}(\text{GP}, \phi)\rightarrow \text{SK}_\phi$.\\
This algorithm generates a private key $\text{SK}_\phi$ for each entity $\phi \in \mathcal{U}_\Phi$.
$\text{SK}_\phi$ is used by a signer $\phi$ to add the signature component to the signcrypted text.

\item $\mathsf{AuthoritySetup}(\text{GP}, \theta)\rightarrow \text{PK}_\theta,\text{SK}_\theta$.\\
This algorithm generates a public/secret key pair $\{\text{PK}_\theta,\text{SK}_\theta\}$ for each attribute authority $\theta \in \mathcal{U}_\Theta$. $\text{PK}_\theta$ is used during the signcryption process, whereas $\text{SK}_\theta$ is used by the authority $\theta$ to generate users' decryption keys.

\item $\mathsf{DecKeyGen}(\text{GID}, \theta , u, \text{SK}_\theta , \text{GP}) \rightarrow \text{DK}_{\text{GID}, u}$.\\
This algorithms takes the global identity of a user $\text{GID} \in \mathcal{GID}$, an attribute $u \in \mathcal{U}_1$, the authority $\theta$ controlling the attribute $u$, the secret key of the authority $\text{SK}_\theta$, and the global parameters. The algorithm outputs $\text{DK}_{\text{GID}, u}$ which is a decryption key for the identity-attribute pair.

\item $\mathsf{VerKeyGen}(\text{GID}, \phi, s, \text{SK}_\phi , \text{GP}) \rightarrow \text{VK}_{\text{GID}, \phi}$.

This algorithms takes the global identity of a user $\text{GID} \in \mathcal{GID}$,
the signer identity attribute $s \in \mathcal{U}_2$ corresponding to the signer entity $\phi$, the signer private key $\text{SK}_\phi$, and the global parameters. The algorithm outputs $\text{VK}_{\text{GID}, \phi}$ which is the key used by user with $\text{GID}$ to verify the messages signcrypted by entity $\phi$.

\item $\mathsf{Signcrypt}(M, (A,\delta), \text{SK}_\phi , \{\text{PK}_\theta\}, \text{GP}) \rightarrow \text{ST}$

This algorithm takes a message $M$, an access structure $(A,\delta)$, the signer private key $\text{SK}_\phi$, a set public keys of the attributes' authorities in the access policy $\{\text{PK}_\theta\}$, and the global parameters $\text{GP}$ and outputs the signcrypted text $\text{ST}$.

\item $\mathsf{Revoke}(\text{ST}, \mathcal{Q}, \{G\}, \text{GP}) \rightarrow \text{ST}'$

This algorithm takes a signcrypted text $\text{ST}$ including its access policy $(A,\delta)$, the set of prime numbers $\mathcal{Q}$, $\{G\}$ which is a set of access lists corresponding to the attributes defining $A$, and the global parameters $\text{GP}$. The algorithm outputs the re-encrypted signcrypted text $\text{ST}'$ such that only users with valid attributes satisfying the access policy can perform designcryption.

\item $\mathsf{Designcrypt}(\text{ST}', \text{VK}_{\text{GID}, \phi}, \{\text{DK}_{\text{GID}, u}\}, \text{GP}) \rightarrow M$.\\
This algorithm takes the re-encrypted signcrypted text $\text{ST}'$, the verification key $\text{VK}_{\text{GID}, \phi}$, the set of decryption keys $\{\text{DK}_{\text{GID}, u}\}$ of a single user with identity $\text{GID}$ corresponding to its attributes, and the global parameters.
The algorithm outputs the message $M$ if and only if the following three conditions are satisfied: (1) the message was signed by $\phi$; (2) the attribute set associated with $\{\text{SK}_{\text{GID}, u}\}$ can satisfy the access policy of the ciphertext, and (3) all the attribute set associated with $\{\text{SK}_{\text{GID}, u}\}$ are valid, i.e., none of them has not been revoked, otherwise, designcryption process fails.

\end{enumerate}

\subsection{System Setup}
\textbf{Generation of Global Parameters}.
At the initial system setup phase, an offline trusted authority (TA) runs the $\mathsf{GlobalSetup}$  algorithm.
First, it defines
the universe of the attributes $\mathcal{U}_1$,
the universe of the attribute authorities $\mathcal{U}_\Theta$,
the universe of the signer entities $\mathcal{U}_\Phi$,
the universe of the signers' identity attributes $\mathcal{U}_2$,
the universe of the global identities $\mathcal{GID}$,
and the mapping function  $\mathsf{T}$$: \mathcal{U}\rightarrow \mathcal{U}_\Psi$, where $\mathcal{U} = \mathcal{U}_1 \cup \mathcal{U}_2$ and $\mathcal{U}_\Psi = \mathcal{U}_\Theta \cup \mathcal{U}_\Phi$.
Then, it generates a bilinear pairing parameters $\left(p, \mathbb{G}, \mathbb{G}_{T}, g, \hat{e}\right)$ where
$\mathbb{G}$, $\mathbb{G}_T$ are a multiplicative cyclic group of prime order $p$,
$g$ is a generator of $\mathbb{G}$,
and $e: \mathbb{G} \times \mathbb{G} \rightarrow \mathbb{G}_T$.
It also chooses two functions, $H$ and $F$, that map the global identities and the attributes to elements in $\mathbb{G}$, respectively, i.e., $H$$:\mathcal{GID}\rightarrow\mathbb{G}$, and $F$$:\mathcal{U}\rightarrow\mathbb{G}$.
Finally, it publishes the global parameters $\text{GP}$ as
$\text{GP}=\{p, \mathbb{G}, g, H, F, \mathcal{U}, \mathcal{U}_\Psi, \mathsf{T}\}$.
Moreover, the TA generates the set of $m$ pairwise relatively prime positive integers $\mathcal{Q}=\{q_1, \dots, q_m\}$ and assigns $q_i$ to a user with $\text{GID}_i$.

\textbf{Setup of Attribute Authorities}.
Each attribute authority $\theta \in \mathcal{U}_\theta$ runs the $\mathsf{AuthSetup}$ algorithm to generate its public/sectet key pair $\{\text{PK}_\theta,\text{SK}_\theta\}$.
The algorithm chooses two random exponents $\alpha_\theta, y_\theta \in \mathbb{Z}_p$ and publishes $\text{PK}_\theta=\{e(g,g)^{\alpha_\theta}, g^{y_\theta}\}$ as the public key of authority $\theta$, whereas the secret key $\text{SK}_\theta=\{\alpha_\theta,y_\theta\}$ is kept secret.

\textbf{Private Key Generation}.
Each signer $\phi \in \mathcal{U}_\Phi$ runs the $\mathsf{SignKeyGen}$ algorithm to generate its private key $\text{SK}_\phi$ that is used to add the signature component to the ciphertext.
The algorithm chooses two random exponents $\alpha_\phi, y_\phi \in \mathbb{Z}_p$ as the private keys that are known only to $\phi$.

\subsection{Users' Key Generation}

Key generation phase consists of two operations; (1) the generation of decryption keys which is executed by attribute authorities and (2) generation of verification keys which is executed by the signers.

\textbf{Generation of Decryption Keys}.
Each attribute authority $\theta$ runs the $\mathsf{DecKeyGen}$ algorithm to generate a decryption key for each identity-attribute pair, i.e., for user with identity $\text{GID}$ holding an attribute $u$. First, the algorithm chooses a random element $t \in \mathbb{Z}_p$.
Then, it computes two components $\text{K}_{\text{GID},u}=g^{\alpha_\theta} H(\text{GID})^{y_\theta} F(u)^t$ and $\text{K}_{\text{GID},u}'=g^{t}$.
Finally, the algorithm outputs the decryption key as $\text{DK}_{\text{GID}, u}=\{\text{K}_{\text{GID},u}, \text{K}_{\text{GID},u}'\}$.

\textbf{Generation of Verification Keys}.
Each signer $\phi$ runs the $\mathsf{VerKeyGen}$ algorithm to generate a verification key for each user. First, the algorithm chooses a random element $r \in \mathbb{Z}_p$. Then, it computes $\text{K}_{\text{GID},\phi}=g^{\alpha_\phi} H(\text{GID})^{y_\phi} F(s)^r$ and $\text{K}_{\text{GID},\phi}'=g^{r}$. Finally, it outputs the verification key as $\text{VK}_{\text{GID}, \phi}=\{\text{K}_{\text{GID},\phi}, \text{K}_{\text{GID},\phi}'\}$.

\subsection{Signcryption} \label{sub:signcryption}
When an signer $\phi$ wants to signcrypt a message $M$, it defines the access matrix $A$
as explained in Appendix \ref{appx_A}.
The boolean formula that should generate the access policy should be on the form
``The signer identity attribute $s \in \mathcal{U}_2$'' \textbf{AND} ``any monotonic boolean formula over $\mathcal{U}_1$''.
It should be the case that $s \in \mathsf{T}^{-1} (\phi)$, i.e., this signer identity attribute $s$ is controlled by the signer $\phi$.

Then, the signcryptor $\phi$ runs the $\mathsf{Singcrypt}$ algorithm to generate the signcrypted text. The algorithm takes a message $M$, an access policy $(A,\delta)$ with $A \in \mathbb{Z}_p^{\ell \times n}$, the public keys of the relevant authorities, the private key $\text{SK}_\phi$, and the global parameters. Let $\rho$ be a mapping function that maps rows from the access policy to attribute authorities, i.e, $\rho$$: \{1, \dots, \ell\} \rightarrow \mathcal{U}_\Theta$ defined as $\rho(.)=\mathsf{T}(\delta(.))$.

First, the algorithm creates vectors $v=(z,v_2,\dots,v_n)^\top$ and $w=(0,w_2,\dots,w_n)^\top$ where $\{z,v_2,\dots,v_n,w_2,\dots,w_n\} \xleftarrow[]{R} \mathbb{Z}_p$. For a secret $z\in\mathbb{Z}_p$, let $\lambda_x$ represents the share corresponding to row $x$ and $w_x$ represents the share of a $0$.
In Appendix \ref{appx_A}, we explain how these shares can be computed.
For each row $x$ of $A$, the algorithm chooses a random $t_x \in \mathbb{Z}_p$ and the signcrypted text is computed as

\begin{equation}
\text{ST}= \
    \begin{pmatrix}
        \begin{split}
            & \ (A,\delta)\ , \ C=Me(g,g)^{z}\ , \  \\
            &\begin{Bmatrix*}[l]
                &C_{1,x}=e(g,g)^{\lambda_x} e(g,g)^{\alpha_{\rho(x)}t_x},\\             & C_{2,x}=g^{-t_x},\\
                &C_{3,x}=g^{y_{\rho(x)}t_x}g^{w_x},\\ &C_{4,x}=F(\delta(x))^{t_x}\\
            \end{Bmatrix*}_{x \in A}\\
        \end{split}
    \end{pmatrix}
\end{equation}

Note that, based on our definition to the boolen formula generating the access structure, there is only one row $x$ for which $\rho(x)=\phi$. Therefore, in order to correctly compute the components  $C_{1,x}=e(g,g)^{\lambda_x} e(g,g)^{\alpha_{\phi}t_x}$ and $C_{3,x}=g^{y_{\phi}t_x}g^{w_x}$, the signcryptor must have the knowledge of $\alpha_{\phi}$ and $y_{\phi}$.
Since these two parameters are the private keys known only to entity $\phi$, no other entity except $\phi$ can compute these components.

\subsection{Revocation}
Before sending the signcrypted text to users, the $\mathsf{Revoke}$ algorithm is used to re-encrypt the signcrypted text such that only users with valid attributes can perform the designcryption process.
For each access list $G_x \in \mathcal{G}$ corresponding to row $x$ in the access policy, the algorithm chooses a random key $\beta_x \in \mathbb{Z}_p^*$ which is a group key for the members of $G_x$ and re-encrypts only one component of the signcrypted text  $C_{2,x}$ to be $C_{2,x}'=C_{2,x}^{\beta_x}=g^{-\beta_x t_x}$.
Only users with valid attribute $\delta(x)$ should be able to recover $\beta_x$ and thus can perform designcryption.
Therefore, for every user with valid attribute, i.e., for every $\text{GID}_i \in G_x$, the algorithm computes $b_i=\beta_x \oplus q_i$ where $q_i$ is the prime number corresponding to $\text{GID}_i$ and $\oplus$ is the XOR operation. Then, the algorithm computes the solution of the CRT congruence system modulo $Q_x=\prod _{i \in G_x} q_i$ as
\begin{equation*}
    B_x=\sum_{i \in G_x} b_iQ_iy_i \mod Q_x
\end{equation*}
and attaches $B_x$ to re-encrypted signcrypted text as follows

\begin{equation}
    \text{ST}'= \
        \begin{pmatrix}
            \begin {split}
                & (A,\delta)\ , \ C=Me(g,g)^{z}\ , \  \\
                & \begin{Bmatrix*}[l]
                        &C_{1,x}=e(g,g)^{\lambda_x} e(g,g)^{\alpha_{\rho(x)}t_x},\\
                        &C_{2,x}'=g^{-\beta_x t_x},\\
                        &C_{3,x}=g^{y_{\rho(x)}t_x}g^{w_x},\\
                        &C_{4,x}=F(\delta(x))^{t_x},\\
                        &B_x\\
                \end{Bmatrix*}_{x \in A}\\
            \end{split}
        \end{pmatrix}
\end{equation}

$B_x$ is used to help the users with valid attributes to recover the secret $\beta_x$ and thus only this set of users can perform decryption as explained in the next subsection.

\subsection{Designcryption}

If a user with identity $\text{GID}$ has a set of valid attributes $S$ that can satisfy the access policy $(A, \delta)$ associated with the signcrypted text and has the verification key of the signer $\phi$, then for each row $x$ corresponding to the attributes in $S$, the user first recovers the group key $\beta_x$ as follows
\begin{equation} \label{eq:decryption_step0}
    \beta_x=(B_x \mod q_i) \oplus q_i
\end{equation}
This is true as the CRT states that $B_x \equiv b_i \mod q_i$ and $b_i \oplus q_i= (\beta_x \oplus q_i) \oplus q_i = \beta_x$.
It is clear that, since the solution of the CRT congruence is constructed using only the prime numbers of users with valid attributes, then only this set of users can reconstruct $\beta_x$ and can proceed with the designcryption process.
Then, the user can recover $C_{2,x}$ from $C_{2,x}'$ as $C_{2,x}=C_{2,x}'^{\frac{1}{\beta_x}}$.
After that, the user computes
\begin{equation} \label{eq:decryption_step1}
	\begin {split}
    D_x
	& =	C_{1,x} \  . \
        e (\text{K}_{\text{GID},\delta(x)},C_{2,x}). 
        e(H(\text{GID}),C_{3,x}) . \\
    & \ \ \ \ e(\text{K}_{\text{GID},\delta(x)}',C_{4,x})\\
	& =	e(g,g)^{\lambda_x} e(H(\text{GID}),g)^{w_x}
	\end {split}
\end{equation}
The correctness proof of \autoref{eq:decryption_step1} is as follows.
\begin{equation}
	\begin {split}
    D_x
	& =	C_{1,x}
        e (\text{K}_{\text{GID},\delta(x)},C_{2,x})
        e(H(\text{GID}),C_{3,x})\\ 
    & \ \ \ \ e(\text{K}_{\text{GID},\delta(x)}',C_{4,x})\\
    & =	e(g,g)^{\lambda_x} e(g,g)^{\alpha_{\rho(x)}t_x}
        e (g^{\alpha_\theta} H(\text{GID})^{y_\theta} F(u)^t,g^{-t_x})\\ 
    & \ \ \ \ e(H(\text{GID}),g^{y_{\rho(x)}t_x}g^{w_x})
        e(g^t,F(\delta(x))^{t_x})\\
    & =	e(g,g)^{\lambda_x} e(g,g)^{\alpha_\theta t_x}
        e (g^{\alpha_\theta} H(\text{GID})^{y_\theta} F(u)^t,g^{-t_x}) \\
    & \ \ \ \ e(H(\text{GID}),g^{y_\theta t_x}g^{w_x})
        e(g^t,F(u)^{t_x})\\
    & =	e(g,g)^{\lambda_x} \ e(g,g)^{\alpha_\theta t_x}
        e (g, g)^{- \alpha_\theta t_x}
        e (H(\text{GID}), g)^{- y_\theta t_x}\\
    & \ \ \ \ e (F(u), g)^{-t t_x}
        e(H(\text{GID}), g)^{y_\theta t_x}
        \\& \ \ \ \ \
        e(H(\text{GID}), g)^{w_x}
        e(g,F(u))^{t t_x}\\
    & =	e(g,g)^{\lambda_x} \ e(H(\text{GID}),g)^{w_x}
	\end {split}
\end{equation}

Then, the user computes
\begin{equation} \label{eq:decryption_step2}
    D = \prod_{x \in S} D_x^{c_x} =	e(g,g)^z
\end{equation}
The correctness proof of \autoref{eq:decryption_step2} is as follows.
\begin{equation} \label{eq:dec_step2_proof}
	\begin {split}
    D
    & = \prod_{x \in S} D_x^{c_x}\\
	& =	\prod_{x \in S} \Big(e(g,g)^{\lambda_x} \ e(H(\text{GID}),g)^{w_x}\Big)^{c_x}\\
	& =	\prod_{x \in S} e(g,g)^{\lambda_x c_x} \ e(H(\text{GID}),g)^{w_x c_x}\\
    & =	e(g,g)^{\sum_{x \in S} \lambda_x c_x} \ e(H(\text{GID}),g)^{\sum_{x \in S} w_x c_x}\\
    & =	e(g,g)^{z} \ e(H(\text{GID}),g)^{0}\\
    & =	e(g,g)^z
	\end {split}
\end{equation}
Finally, the user can recover the message as
\begin{equation}
    \frac{C_0}{D} = \frac{M \ e(g,g)^z}{e(g,g)^z} = M	
\end{equation}

\subsection{Using our scheme in multi-authority AMI networks}
The aforementioned design goals can be achieved using our attribute-based signcryption scheme by mapping the multi-authority AMI network entities to our scheme as follows.
The DNOs' set $\mathbb{D}$ and the vendors' set $\mathbb{V}$ are mapped to the universe of attribute authorities $\mathcal{U}_\Theta$.
This is because each DNO and vendor should be able to issue different attributes for their customers such as location attribute, electricity plan attribute, DLC membership attribute, etc.
Also, the same sets are mapped to the universe of signer entities $\mathcal{U}_\Phi$.
This is because DNOs and vendors need to send authenticated multicast messages to their customers using signcryption.
The users' set $\mathbb{U}$ is mapped to the universe of global identifiers $\mathcal{GID}$.
Upon registration, each user $U_i$ should receive its unique prime number $q_i$, decryption keys and a verification key from the DNOs and vendors.
In order to send a multicast message, a DNO or vendor should define the monotonic boolean relation under which a message is signcrypted and call the $\mathsf{Signcrypt}$ algorithm.
Then, the signcrypted text is broadcasted to all users through the DCC and the hierarchal network structure.
According to \cite{DEP2SA} and \cite{EPDA}, the DCC is the entity that manages the supplier-users relationship, i.e., the DCC by default learns the set of attributes of each user, but it does not know their decryption keys. Therefore, the DCC is the entity that can run the $\mathsf{Revoke}$ algorithm to ensure that users with revoked attributes cannot decrypt the multicasted messages.
Finally, upon receiving a multicast message, the user checks if his non-revoked attributes can satisfy the access policy, then he calls the $\mathsf{Designcrypt}$ algorithm to decrypt the message, otherwise, the message cannot be decrypted and should be discarded.

\section{Security Analysis} \label{sec:multicast_security_analysis}

\subsection{Collusion Resistance}
In collusion attack, several users may collude by combining their attributes to satisfy the access policy of a ciphertext and decrypt it, i.e., they combine their decryption keys to run the $\mathsf{designcrypt}$ algorithm to decrypt a signcrypted text which they cannot decrypt individually.
This attack cannot succeed in our scheme.
During the designcryption process the shares of the ``0'', $w_x$ values, are crucially engaged to the global identifier of the secret key of the user as in \cite{RW2015_MA_ABE} and \cite{LW2011_MA_ABE}. This is clear in \autoref{eq:dec_step2_proof} where the term $\prod  e(H(\text{GID}),g)^{w_x c_x}$ can be reduced to $ e(H(\text{GID}),g)^{\sum_{x \in S} w_x c_x}=e(H(\text{GID}),g)^{0}$ if and only if a single $\text{GID}$ is used.
Therefore, in case that two or more users collude and try to decrypt the same signcrypted text, the ``0-shares'' will result in a failed decryption, which can thwart collusion attacks.

\subsection{Signature Forgery Resistance and Non-repudiation}
In forgery attack, an adversary $\mathscr{A}$ may try to forge the singcryption of a signer $\phi$.
As discussed in \autoref{sub:signcryption}, computing a valid signature of the signer $\phi$ requires the knowledge of $\alpha_{\phi}$ and $y_{\phi}$ to correctly compute the components $e(g,g)^{\alpha_{\phi}}$ and $g^{y_{\phi}}$.
$\mathscr{A}$ cannot obtain $\alpha_{\phi}$ and $y_{\phi}$ from any verification key $\text{K}_{\text{GID},\phi}=g^{\alpha_\phi} H(\text{GID})^{y_\phi} F(s)^r$ as this requires $\mathscr{A}$ to split the three components and solve the discrete logarithmic problem (DLP) for each component which is infeasible. Therefore, our scheme can resist forging signatures and thus ensure sender authentication and message non-repudiation since entity $\phi$ is the only entity that can compute a valid signature.

\section{Performance Evaluation} \label{sec:multicast_performance}
\begin{figure}[!t]
	\centering
	\includegraphics[clip=true,width=0.4 \textwidth]{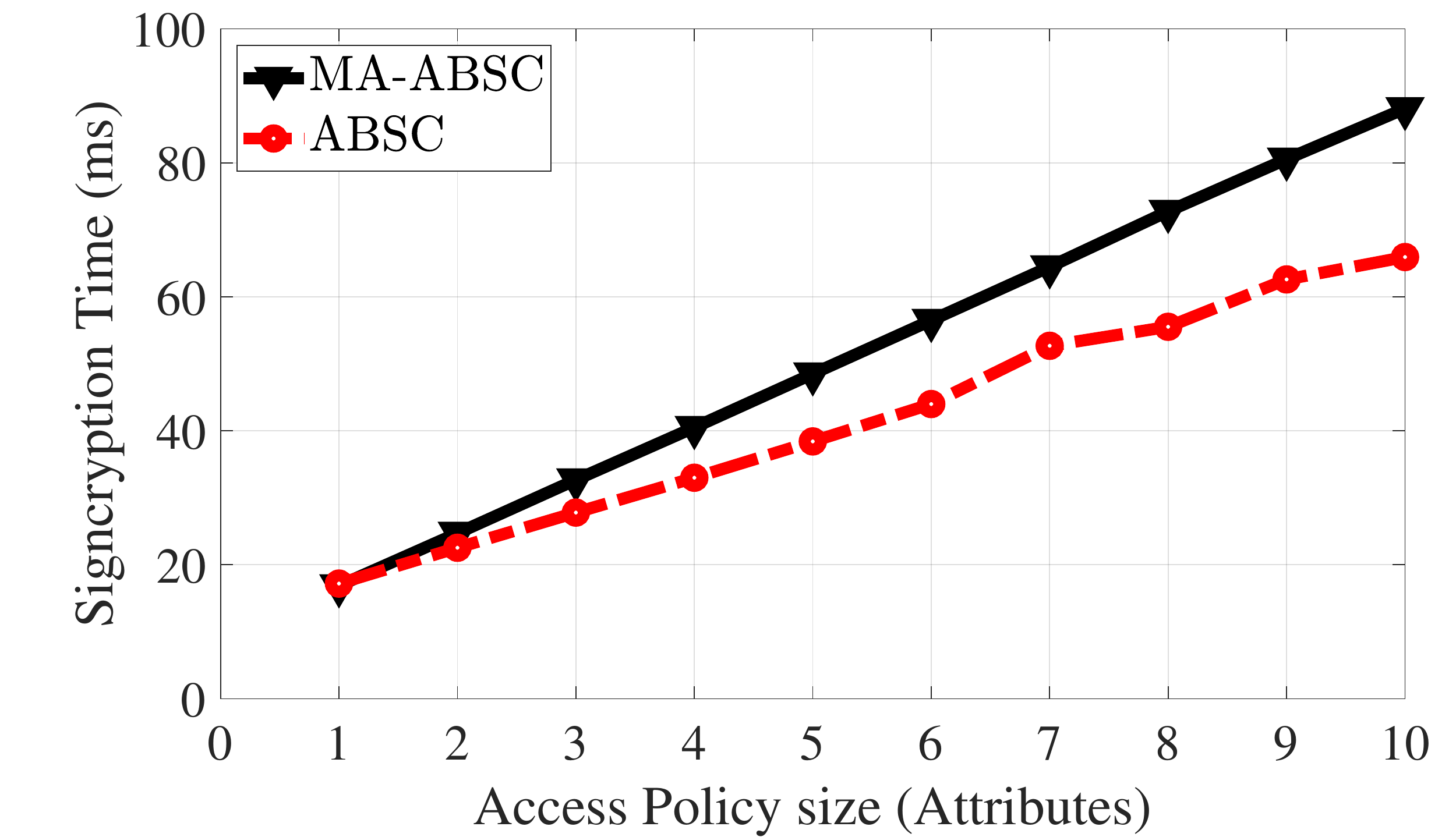}
	\caption{Signcryption time vs access policy size.}
	\label{fig:multicast_signcryption}
    \vspace{-5mm}
\end{figure}

In this section, we compare our multi-authority attribute-based signcryption (MA-ABSC) scheme to the closest similar scheme in \cite{ABSC2015} which is a single-authority attribute-based signcryption scheme (ABSC).
We implemented both schemes using Python charm cryptographic library \cite{charm}.
Supersingular  elliptic  curve  with  the  symmetric  Type  1 pairing of size 512 bits (SS512 curve) is used for all pairing operations.
All  cryptographic  operations  were  run  1,000 times  and  average  measurements  are  reported.
Typically, DNOs, vendors and the DCC have powerful computational resources.
Therefore, in our experiments, they are implemented by a workstation with Intel Core i7-4765T 2.00 GHz and 8 GB RAM. The operations done by the DNOs and vendors are the signcryption processes, whereas the revocation process is executed by the DCC.
On the other hand, to implement the resource-limited SMs, we used Tennessee Tech. University AMI testbed of 30 \textit{Raspberry-Pi} 3 devices with an ARM Cortex-A53, 1.2 GHz processor and 1 GB RAM. The operation executed by the SMs is the designcryption.

\begin{figure}[!t]
	\centering
	\includegraphics[clip=true,width=0.4 \textwidth]{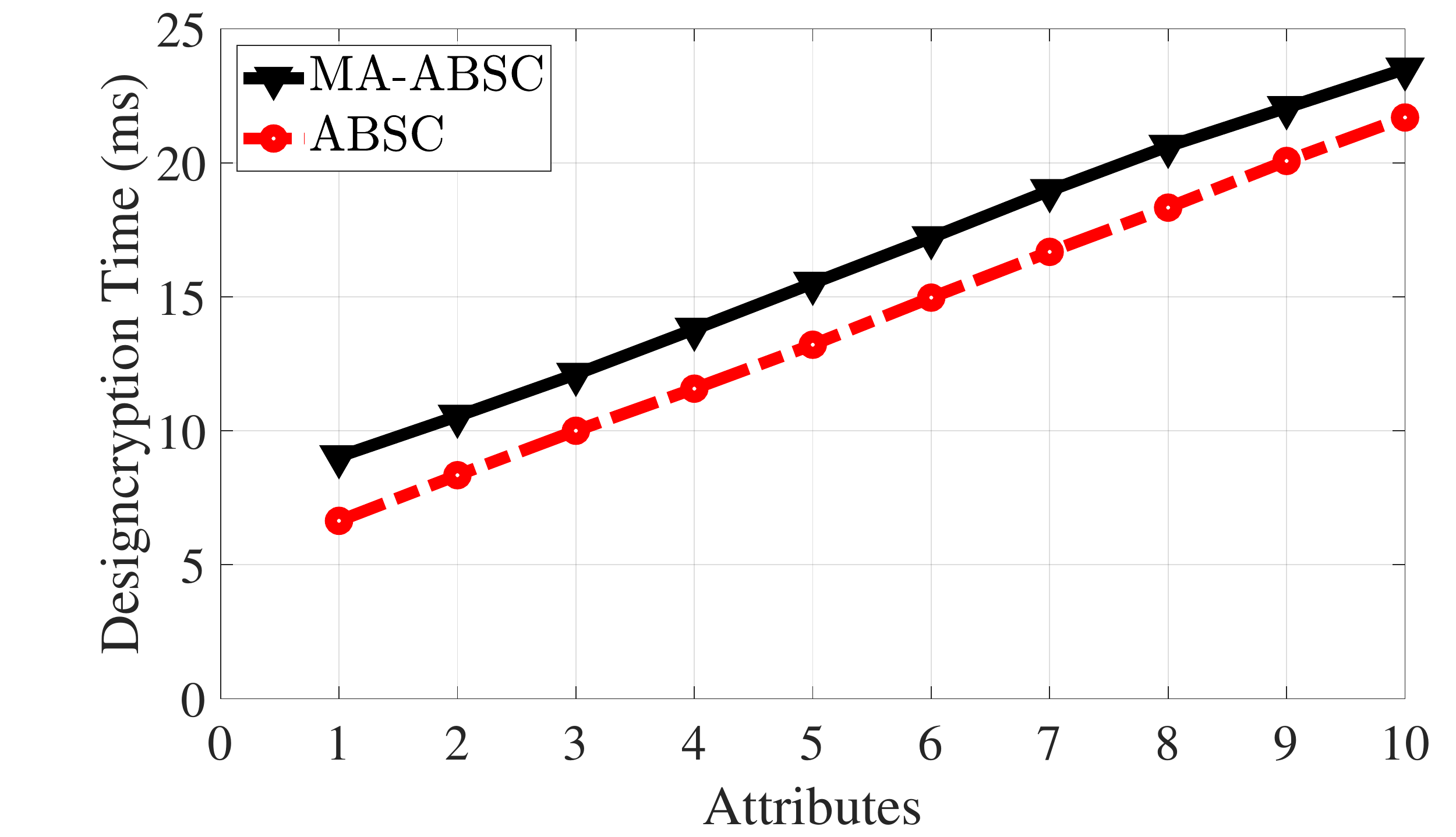}
	\caption{Deigncryption time vs access policy size.}
	\label{fig:multicast_designcryption}
    \vspace{-5mm}
\end{figure}

\begin{figure}[!t]
	\centering
	\includegraphics[clip=true,width=0.4 \textwidth]{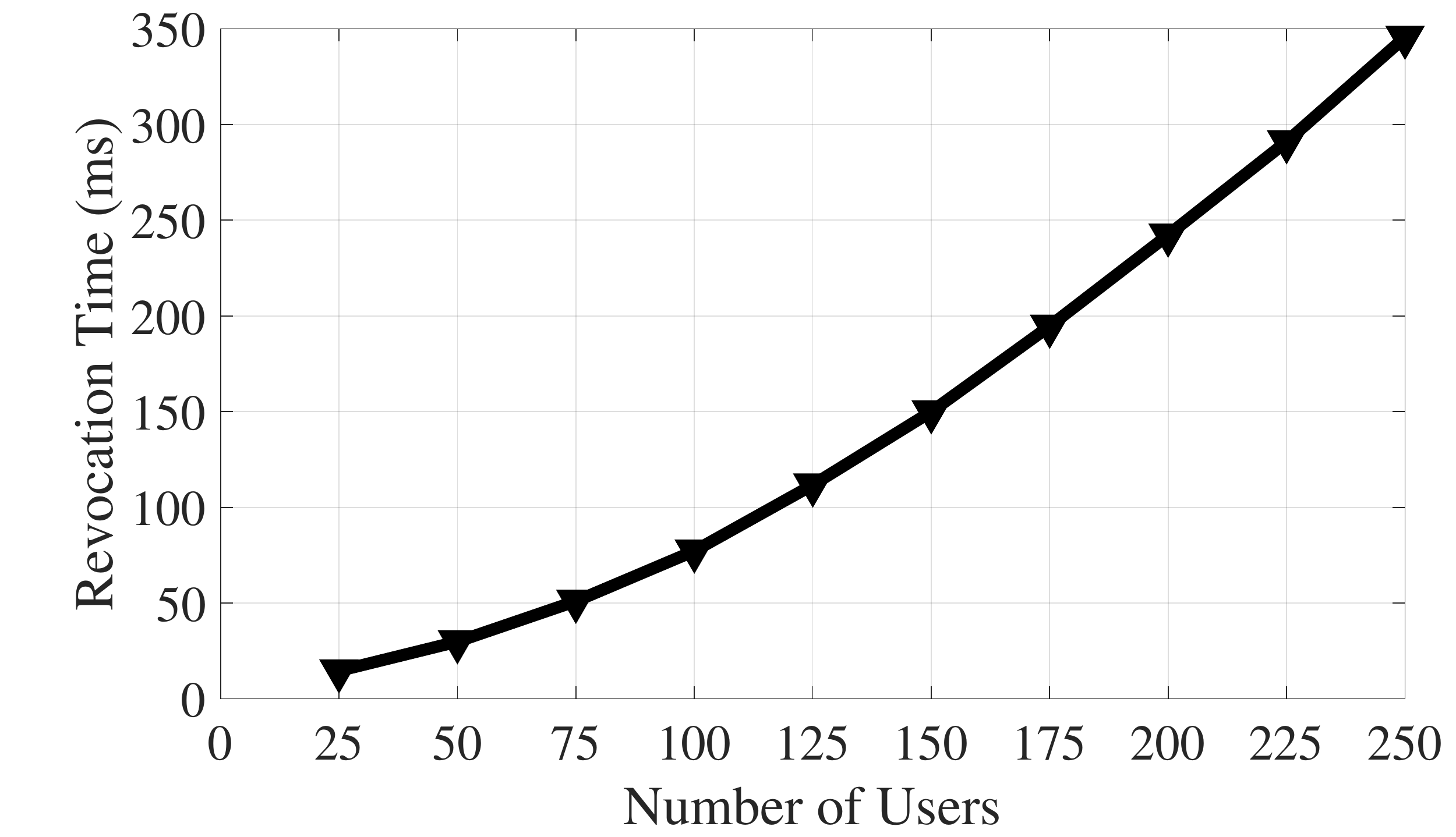}
	\caption{Revocation time vs number of users.}
	\label{fig:revoc}
    \vspace{-5mm}
\end{figure}

Figure \ref{fig:multicast_signcryption} gives the signcryption time versus the access policy size used to signcrypt a message.
As shown in the figure, our scheme has slightly higher signcryption time than the scheme in \cite{ABSC2015}.
Figure \ref{fig:multicast_designcryption} gives the designcryption time versus the number of attributes used during the designcryption process.
As shown in the figure, the designcryption time is close to that of  \cite{ABSC2015}.
As compared to ABSC \cite{ABSC2015}, the increased computation cost in the signcryption and designcryption processes are needed to allow multiple authorities to control their own attributes which cannot be achieved in \cite{ABSC2015} that allows only a single authority to control the whole attribute set.

Lastly, we plot in \autoref{fig:revoc} the revocation computation cost of our scheme as the number of users with valid attributes increases.
ABSC \cite{ABSC2015} is not considered in this evaluation since it does not support attribute revocation.
As shown in the figure, the revocation process adds an acceptable cost to our multicast scheme and the times are in the range of milliseconds. For instance, the revocation computation cost is only 0.34 second for a case in which the access list contains 250 users with valid attribute.
To conclude, compared to the baseline attribute-based signcryption scheme in \cite{ABSC2015}, our scheme achieves more features with acceptable additional computation cost.

\section{Conclusions} \label{sec:multicast_conclusions}

    In this paper, we proposed an attribute-based signcryption scheme that can be used to secure SG downlink multicast communication.
    The proposed scheme can achieve data confidentiality, message source authentication, message non-repudiation, and immediate attribute revocation, simultaneously which are required for secure multicast communications.
    In addition, the scheme can resist collusion attacks in which several users collude to decrypt a ciphertext they cannot decrypt individually.
    Our security analysis confirms that the proposed scheme is secure and can achieve the aforementioned features.
    Our experiments conducted on the AMI testbed at Tennessee Tech. University confirms that the proposed scheme has low values of computational overheads which is required for resource-constrained SMs.

\appendices
\section{Generating LSS Matrices from Monotonic Boolean Formulas} \label{appx_A}

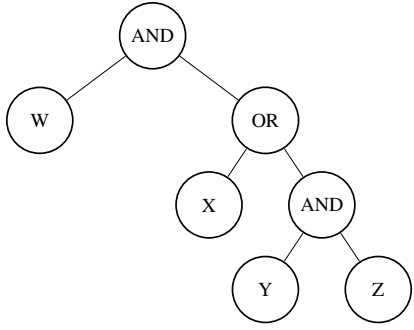
\begin{figure}
    \centering
    \tikzstyle{circle1} = [circle, draw, fill=white, text width=2.4em, text centered, rounded corners, minimum height=1.5em, line width=0.3mm]
    \scalebox{0.75}
    {
        \begin{tikzpicture}[level distance=1.5cm,
          level 1/.style={sibling distance=4cm},
          level 2/.style={sibling distance=2cm}] 
          \node[circle1] {AND}
            child
            {
                node[circle1] {W}
            }
            child
            {
                node[circle1] {OR}
                    child
                    {
                        node[circle1] {X}
                    }
                    child
                    {
                        node[circle1] {AND}
                            child
                            {
                                node[circle1] {Y}
                            }
                            child
                            {
                                node[circle1] {Z}
                            }
                    }
            };
        \end{tikzpicture}
    }
    \caption{Access Tree Example.}
    \label{fig:access_tree}
    \vspace{-5mm}
\end{figure}

A monotonic boolean formula can be represented as a binary access tree in which interior nodes are AND and OR gates while the leaf nodes represent attributes. \autoref{fig:access_tree} shows the access tree for the boolean formula $\text{W AND } \big(\text{X OR }(\text{Y  AND Z})\big)$ where W, X, Y, and Z are the attributes.

According to \cite{LW2011_MA_ABE}, the following algorithm can convert a monotonic boolean formula into an equivalent LSS matrix.
First, a counter $c$ is initialized by one and the tree root node is labeled with a vector $(1)$ (a vector of length $c$).
Then, each child node is labeled with a vector determined by the vector assigned to its parent node as follows.
If the parent node is an OR gate labeled by the vector $v$, then its children are labeled by $v$ (and the value of $c$ stays the same).
If the parent node is an AND gate labeled by $v$, first $v$ is padded with zeros at the end (if necessary) to make it of length $c$.
Then, one child node is labeled with the vector $v|1$ (where $|$ denotes appending a new element to vector $v$) and the other child node is labeled with the vector $(0, \dots, 0)|-1$, where $(0, \dots, 0)$ denotes a zero vector of length $c$. Note that the summation of these two vectors is $v|0$. Finally, $c$ is incremented by one.
The process continues in a top-bottom manner until all the entire tree nodes are labeled.
Once the entire tree is labeled, the vectors labeling the leaf nodes form the rows of the LSS matrix.
If these vectors have different lengths, the shorter vectors are padded with zeros at the end.

\begin{figure}
    \centering
    \tikzstyle{circle1} = [circle, draw, fill=white, text width=2.4em, text centered, rounded corners, minimum height=1.5em, line width=0.3mm]
    \scalebox{0.75}
    {
        \begin{tikzpicture}[level distance=1.5cm,
          level 1/.style={sibling distance=4cm},
          level 2/.style={sibling distance=2cm}] 
          \node[circle1] {1}
            child
            {
                node[circle1] {1,1}
            }
            child
            {
                node[circle1] {0,-1}
                    child
                    {
                        node[circle1] {0,-1}
                    }
                    child
                    {
                        node[circle1] {0,-1}
                            child
                            {
                                node[circle1] {0,-1,1}
                            }
                            child
                            {
                                node[circle1] {0,0,-1}
                            }
                    }
            };
        \end{tikzpicture}
    }
    \caption{Access Tree with Labels.}
    \label{fig:access_tree_labeled}
    \vspace{-5mm}
\end{figure}
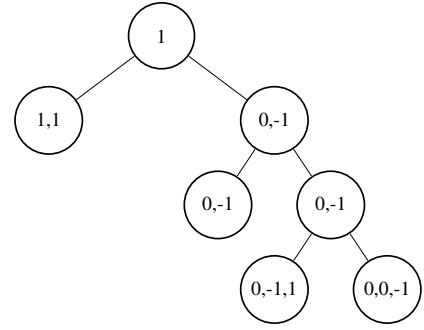

For the tree shown in \autoref{fig:access_tree}, the root AND node is labeled (1), its left child, node (W) node, is labeled (1, 1) while its right child, node (OR), is labeled by $(0,-1)$.
Then, both children of the OR node, nodes (X) and (AND), are labeled $(0,-1)$ as their parent.
Finally, the left child of the AND node, node (Y), is labeled $(0,-1,1)$ while the right child, node (Z), is labeled $(0,-1,-1)$. \autoref{fig:access_tree_labeled} shows the fully labeled tree.
The resulting LSS matrix after padding leaf nodes is

\begin{equation*}
    A=
    \begin{pmatrix}
        1 & 1  & 0  \\
        0 & -1 & 0  \\
        0 & -1 & 1  \\
        0 & 0  & -1 \\
    \end{pmatrix}
\end{equation*}

To generate the secret shares, let a secret $z=65$, construct the column vector $\mathbolditalic{v} = (z, r_2, \dots , r_n)= (65, 3, 4)$ where $3, 4$ are random numbers, then compute the shares vector $\boldsymbol \lambda = A\mathbolditalic{v}$ as
\begin{equation*}
    \boldsymbol \lambda=A\mathbolditalic{v}=
    \begin{pmatrix}
        1 & 1  & 0  \\
        0 & -1 & 0  \\
        0 & -1 & 1  \\
        0 & 0  & -1 \\
    \end{pmatrix}
    \begin{pmatrix}
        65  \\
        3  \\
        4  \\
    \end{pmatrix}
    =
    \begin{pmatrix}
        68  \\
        -3  \\
        1   \\
        -4  \\
    \end{pmatrix}
\end{equation*}

To reconstruct the secret, we recall $\sum_{i \in I} c_i \lambda_i=z$. However, the aforementioned algorithm forces the reconstruction coefficients $\{c_i\}_{i \in I}$ to have a value of one. This means that adding the secret shares corresponding to attributes validating the boolean formula can reconstruct the secret. For example (W AND X) can satisfy the boolean formula, therefore, adding the shares corresponding to W, which is $68$, to the share corresponding to X, which is $-3$, can reconstruct the secret $65$. The same applies to (W AND Y AND Z).

\bibliographystyle{IEEEtran}
\bibliography{Paper}

\end{document}